# On Invariance and Convergence in Time Complexity theory.

Mircea Alexandru Popescu Moscu
Bucharest, Romania
mircalex@yahoo.com

**Astract**

The present paper intends to introduce a principle of invariance of "complexity hierarchies" under any "reasonable" changing of the input's dimension measure function. For a Turing machine that takes as input a word of a given length (say n) the input's dimension is n, so in fact the "dimension measure" is the identity function. We state that the set inclusion relation between the complexity class C and it's nondeterministic extension class NC is invariant under any monotonic crescent injective measure function. Apart from this there comes another statement of invariance saying that the inclusion relation is kept even if any set of Turing machine transitions, determined by some general property in all conceivable Turing machines, are not counted when determining the computation's length. As an example, we make reference to the set of transitions that move the head left or right without reading or writting -- in a computer program accesing the first or the last element of an array is usually independent of the size of the array, i.e is constant, thus not depending on the size of the array ( moving through the array is not counted ). We must state that this paper was developed thinking only time complexity, although we see no reason why it couldn't be ported to space complexity also. Finally, we'll prove that for any language there is an infinite sequence of languages from O(n) that converges to that language.

## 1. Definitions

### 1.1 The set of complexity measure functions

**CMF =** { f: $(0,\infty) \to (0,\infty)$ | $\forall$ x,y $\in$ $(0,\infty)$ ( x < y ) $\Rightarrow$ ( f(x) < f(y)) }
As can be seen this is the set of positive injective crescent functions on $(0,\infty)$

### 1.2 The complexity class determined by a subset of CMF

Let there be a set of functions **S ⊆ CMF**. We define the complexity class associated with **S** (denoted by **C(S)**) as follows: a language L, over an alphabet Σ, is in **C(S) iff** ∃ M a Turing machine that accepts L and ∃ f ∈ **S** such that on any x ∈ Σ* M stops in at most f(|x|) steps (where |x| denotes the length of the word x)

### 1.3 Nondeterministic extension of a class

For a complexity class **C=C(S)** we define the nodeterministic extension **NC** as the class of languages that have a checking relation in the class C. A more formal definition should be like this: a language L over Σ is in **NC** iff there is a checking relation R in **C** and a function g in **S** such that for all x ∈ Σ*,
x ∈ L ⇔ ∃y( |y| ≤ g(|x|) and (x,y) ∈ R )

### 1.4 Complexity class transformation

Let there be a complexity class **C=C(S)** for some **S ⊆ CMF** and a function T ∈ **CMF**. The complexity class **T(C)** is called the transformation of **C** through **T** and is defined bearing in mind that **C** is the complexity class determined by the set of complexity measuring functions **S**: a language L over Σ is in **T(C)** iff there exists a Turing machine M that accepts L and a function g ∈ **S** such that for any x ∈ Σ* M stops in at most g(T(|x|)) steps.

Obs: a similar definition could be given by imposing that M stops in T(g(|x|)) steps.

### 1.5 Complexity class inverse-transformation

Similar in everything to the definition **1.4** only that we require M to stop in T(g(|x|)) steps (see observation in previous definition). We will denote this complexity class by $T_{INV}$(**C**).

### 1.6 Nondeterministic extension of a complexity class transformation.

For a complexity class **C=C(S)** and a function **T ∈ CMF** we define the nodeterministic extension **NT(C)** as the class of languages that have a checking relation in the class **T(C)**. A more formal definition should be like this: a language L over Σ is in **NT(C)** iff there is a checking relation R in **T(C)** and a function g in **S** such that for all x ∈ Σ*,
x ∈ L ⇔ ∃y( T(|y|) ≤ g(T(|x|) ) and (x,y) ∈ R )

## 2. Invariance principle of complexity class transformation

**2.1 Direct Principle**: For any complexity class **C=C(S)** where S is an infinite nonempty set, (C = NC) ⇔ (for any T ∈ **CMF** **T**(C)= **NT**(C))

**2.2 Inverse Principle**: For any complexity class **C=C(S)** where S is an infinite nonempty set, (C = NC) ⇔ (for any T ∈ **CMF** **T**$_{INV}$(C)= **NT**$_{INV}$(NC))

Obs: The requirement in the definition of **CMF** (definition **1.1**) to use only injective functions eliminates any concerns arrising from the fact that the class of constant complexity problems seems to be equal with it's nondeterministic extension

## 3. Selection of countable transititions

Suppose we have a function ( let's call it the selection function ) that, for any given Turing machine M = ($\Sigma, \Gamma, Q, \delta$) can tell us which transitions should be counted and which not.
It is obvious that in such conditions we could redefine the lenght of a computation summing up only the transitions indicated by the selection function.
For a selection function f and for a complexity class C=C(S) we'll denote by f(C) the complexity class obtained by changing the counting of length computation as required by the function f.

**3.1 Principle of invariance under selection functions**: For any selection function f and for any complexity class C=C(S), (C = NC) ⇔ (f(C)=Nf(C))

Obs: The principles 2.1 and 2.2 state the invariance of nondeterministic extensions under changes in measuring the input and the principle 3.1 under changes in measuring the computation's length. So we have two different types of invariance: under changes of input measure and under changes of time measure. The principle 2.2 can be viewed as being related to 3.1.

**3.2 Implications of these principles.**

a) If these three principles hold good than we have **LOGTIME=NLOGTIME iff P=NP**. (from principle 2.1 under log and exp transformations )
b) But under a selection function that doesn't count the moving of the head (or in a program where the accesing of a vector cell is considered to be done in constant time no mater how big the vector or how far from the beginning the cell), the following language is in NLOGTIME and is not in LOGTIME:
The language over the alphabet {0,1} consisting of words in which occurs at least once the letter "1". This language has a checking relation which is O(1) (given a word and an index in that word we only need to acces the letter at that index). The algorithm recognizing a word must check, in the worst case, all the cells, thus being lower bounded by n. So, **LOGTIME ≠ NLOGTIME**

Summing up a) and b) we obtain **P ≠ NP**

**4. Convergence of languages.**

    **4.1 Defition.** An infinite sequence of languages $L_n$ over an alphabet $\Sigma$, $n \in N$ converges to a language L over $\Sigma$ iff $\forall k \in N$ $\exists n_k \in N$ and $\forall n \in N$ $n \geq n_k \Rightarrow$ ( $\forall w \in \Sigma^*$ if $|w| \leq |k| \Rightarrow (w \in L_n \Leftrightarrow w \in L)$ )

More intelligibly, a sequence of languages converges to a language if for any natural number **k** there exists another natural number **m** such that all languages in the sequence, having an index greater than **m**, contain exactly the same words as the limit language, provided that we inspect only the words with lengths that are lower or equal to **k**.

    **4.2 Convergence theorema.** For any language L over an alphabet $\Sigma$ there exists a sequence of languages in O(n) that converges to L.

    **Proof:**
Let $L_n$ be the language over $\Sigma$ accepted by the Turing machine $TM_n$. And now we will construct this macchine.
We confine ourselves only to languages over the alphabet {0,1} since the generality of the proof is not affected.

$TM_n = (\Sigma, \Gamma, Q_n, \delta_n)$.
$\Sigma = \{0,1\}$
$\Gamma = \Sigma \cup \{b\}$
$Q_n = \{q_{start}, q_{accept}, q_{reject}, q_0, q_2, \ldots, q_{2^n-1}\}$
$\delta_n: (Q_n - \{q_{start}, q_{accept}\}) \times \Gamma \to Q_n \times \Gamma \times \{-1, 1\}$

if $\lambda \in L$ (the word with no letters, nil word)
  $\delta_n(q_{start}, b) = (q_{accept}, b, 1)$
else
  $\delta_n(q_{start}, b) = (q_{reject}, b, 1)$

We consider that the words over $\{0,1\}$ are natural numbers represented in base 2. We consider that these representations of the numbers are arranged backwards on the tape.

$\delta_n(q_{start}, 0) = (q_0, b, 1)$
$\delta_n(q_{start}, 1) = (q_1, b, 1)$

for any k from 0 to $2^n-1$, let $(k_i)$ with i from 0 to $\lceil \log_2(k) \rceil - 1$ be the representation of k in base 2.

for any k from 0 to $2^{n-1}-1$ and for any $x \in \{0,1\}$ $\delta_n(q_k, x) = (q_p, b, 1)$, where p has the representation $(p_i)$ with i from 0 to $\lceil \log_2(k) \rceil$ and $p_i = k_i$ for all i from 0 to $\lceil \log_2(k) \rceil - 1$ and $q_h = x$, where $h = \lceil \log_2(k) \rceil$

for any k from $2^{n-1}$ to $2^n-1$ and for any $x \in \{0,1\}$ $\delta_n(q_k, x) = (q_{reject}, b, 1)$

for any k from 0 to $2^n-1$, if $w \in L$, where $w = (p_i)$ and i goes from 0 to $z = \lceil \log_2(k) \rceil - 1$ and $p_i = k_{z-i}$, then $\delta_n(q_k, b) = (q_{accept}, b, 1)$, else $\delta_n(q_k, b) = (q_{reject}, b, 1)$

It is obvious that $L_n = L \cap \Sigma_n$ where
$\Sigma_n = \{w \mid w \in \Sigma^* \text{ and } |w| \leq n\}$ and hence the sequence $L_n$ converges to L.

It is obvios that any computation requires only n+1 steps to complete, n being the length of the input word (not counting the blank symbol).

q.e.d

## 5. Conclusion

It seems that P is not equal with NP, if we're to believe that the principles on invariance hold good, and that the property of a complexity class to be closed or opened to the nondeterministic extension operator it's an invariant of complexity theory.

We have proven that any language can be approximated how well we desire by a sequence of liniar time ( O(n) ) languages. But this has non practical importance: first because the memory space required is exponential, and secondly because the constructivity of such a sequence of languages may be impossible (the proof of the existance of such a sequence uses the "Axiom of choice" ). After we've constructed a database of results of a certain magnitude for a problem, solving it is only a question of searching in the database.

All this being said and equal, we rest.